\documentclass[aps,preprint,amsmath,amssymb,amsfonts,nofootinbib]{revtex4-1}
\usepackage{epsfig}
\usepackage{graphicx}
\usepackage{dcolumn}
\usepackage{bm}
\usepackage{amsthm}
\usepackage{amsmath}
\usepackage{color}
\newcommand{\beq}{\begin{equation}}
\newcommand{\eeq}{\end{equation}}
\newcommand{\bea}{\begin{eqnarray}}
\newcommand{\eea}{\end{eqnarray}}

\begin{document}

\title{Topology and the Conformal Invariance of Nodal Lines in Two-Dimensional Active Scalar Turbulence}
\author{Christopher Eling}
\email{cteling@gmail.com}
\noaffiliation

\begin{abstract}

The inverse cascade in two-dimensional hydrodynamic turbulence exhibits a mysterious phenomenon.  Numerical simulations have shown that the nodal isolines of certain scalars actively transported in the flow (eg, the vorticity in Navier-Stokes theory) obey Schramm-Loewner evolution (SLE), which indicates the presence of conformal invariance. Therefore, these turbulent isolines are somehow in the same class as cluster boundaries in equilibrium statistical mechanical models at criticality, such as critical percolation. In this paper, we propose that the inverse cascade is characterized by a local energy (or in some cases, enstrophy) flux field that spontaneously breaks time reversal invariance.  The turbulent state consists of random constant flux domains, with the nodal isolines acting as domain walls where the local flux vanishes.  The generalized circulation of the domains is proportional to a topological winding number.  We argue that these turbulent states are gapped states, in analogy with quantum Hall systems.  The turbulent flow consists of many strongly coupled vortices that are analogous to quasi-particles.  The nodal isolines are associated with the gapless topological degrees of freedom in the flow, where scale invariance is enhanced to conformal invariance.  We introduce a concrete model of this behavior using a two-dimensional effective theory involving the canonical Clebsch scalars.  This theory has patch solutions that exhibit power law scaling. The fractional winding number associated with the patches can be related to the Kolmogorov-Kraichnan scaling dimension of the corresponding fluid theory.  We argue that the fully developed inverse cascade is a scale invariant gas of these patches.  This theory has a conformally invariant sector described by a Liouville conformal field theory whose central charge is fixed by the fractional winding number.  The nodal isolines are defect lines in this theory.

\end{abstract}

\maketitle

\section{Introduction}

Hydrodynamic turbulence occurs when the velocity and physical dimensions of a fluid flow are such that inertial forces dominate over viscous forces.  As the damping and smoothing effect of viscosity is reduced, flows become unstable, leading to highly irregular and chaotic behavior.  Many aspects of turbulent flows remain poorly understood, but there is evidence that general principles such as symmetry and anomalies are an important part of the solution.

One can treat turbulent flows statistically by introducing a random forcing term in the fluid equation (eg, the incompressible Navier-Stokes equation), which creates a stochastic partial differential equation.   The flows can then be characterized by two-(or higher) point correlation functions of fields (eg, velocity, vorticity, etc).  One would like to consider a steady state where the statistics and the correlation functions are time independent.  However, unlike the Langevin-type equations, which describe fluctuations around equilibrium, the fluid equations are non-linear and the turbulent state is far from equilibrium.

Turbulent flows exhibit a cascade, where a conserved quantity such as energy is transferred across widely separated scales and is not in equipartition.  In a direct cascade, the force term injects energy, which is subsequently transferred to smaller length scales until it is dissipated by viscosity, which acts as a UV cutoff.  As viscosity goes to zero and the cutoff is removed, the cascading energy flux remains non-zero and time reversal symmetry is not restored.  This dissipative anomaly is due to divergences in velocity gradients as viscosity goes to zero.

In 1941 Kolmogorov constructed a theory of this fully developed turbulence using symmetry principles \cite{Frischbook}.  Assuming homogeneity, isotropy, and scale invariance, the probability density function $P(\delta v^{\parallel}, r)$ of longitudinal velocity differences $\delta v^{\parallel}$ at separation distances $r$ much greater than the scale of viscous dissipation, but much less than the forcing scale has the form\footnote{Longitudinal differences at scale $r$ are defined as $\delta v^{\parallel} = (\vec{v}(\vec{x}+\vec{r})-\vec{v}(\vec{x})) \cdot \hat{r}$}
\begin{align}
P(\delta v,r) = \frac{1}{\delta v^{\parallel}} f\left(\frac{\delta v^{\parallel}}{r^\Delta}\right).
\end{align}
If the energy flux in the cascade is assumed to be constant, then dimensional analysis yields the famous scaling law $\Delta = -1/3$.  However, there is a significant amount of numerical and experimental evidence that scale invariance is broken and the Kolmogorov scaling is invalid.

It is also interesting to consider turbulent flows in two-dimensions (2d).  2d flows are physically relevant as models of planet atmospheres and oceans on large scales.  Although it might seem that the flows would be simpler,  Kraichnan \cite{Kraichnan} discovered that there can be a double cascade due to the presence of an additional conserved quadratic variable in ideal flows, the enstrophy, which is related to square of the 2d vorticity scalar.  While enstrophy cascades directly to smaller scales, energy is transferred to larger scales in an ``inverse cascade''.  For the inverse cascade, the dissipative anomaly in energy is not present because the Navier-Stokes equations imply that the enstrophy must decrease with with time.  Hence, as viscosity as goes zero, there is no compensating divergence in squared vorticity.  Numerical simulations have indicated that scale invariance is not broken and the turbulent flow exhibits some self-similarity.  The Kolmogorov-Kraichnan scaling appears to hold in the inertial range, i.e. for distances greater than the forcing scale, but smaller than the scale of dissipation from a large-scale friction\footnote{However, see \cite{Mueller} for some recent evidence of intermittency in some statistics in 2d}.

It is tempting to postulate that the symmetries of 2d turbulence can be extended from scale invariance to conformal invariance.  In 2d, conformal invariance is infinite dimensional and the resulting conformal field theories (CFT) are parametrized by the Virasoro central charge $c$ \cite{DiFrancesco:1997nk}.  In the 1990's Polyakov \cite{Polyakov:1992er} attempted to develop a CFT description of turbulence, focusing on the direct enstrophy cascade.   For $c \leq 0$ 2d CFT's are non-unitary, which could account for the enstrophy flux.  However, apparently it is not possible for these models to account for all aspects of turbulence \cite{Falkovich:1993fg}.

On the other hand, in 2006 it was shown numerically that zero vorticity nodal isolines in the inverse cascade are identical to the curves forming cluster boundaries in the scaling limit of statistical mechanical systems at equilibrium \cite{conformal1}.  The nodal curves are expected to be fractal (or possibly multi-fractal) in turbulence, but it is surprising that the measure on the curves is conformal and described by Schramm-Loewner evolution (SLE) \cite{Schramm:1999rp}.  SLE curves have been proven to be the scaling limits of cluster boundaries formed by various statistical mechanical models defined on a lattice (e.g. the Ising model). For a review, see \cite{Bauer:2006gr}.  At criticality and in the continuum limit, these models are described by different 2d CFT's.  Vorticity nodal lines are in the same universality class as critical percolation.  Later work extended these results to turbulence of 2d ``active scalar'' models that are generalizations of the Euler/Navier-Stokes theory.  The zero isolines of a generalized scalar correspond to SLE curves associated with different statistical models \cite{conformal2, Falkovich}.

The equivalence of the nodal lines in 2d turbulence and critical curves in statistical mechanical models is unexpected and mysterious.  An initial thought is that perhaps the curves can be mapped into the isoheight lines of a Gaussian random field.  However, vorticity cannot be Gaussian in turbulence because it has a non-zero three-point function \cite{FalkovichRev}.  Somehow the turbulent flow is able to create curves that have the same statistics as the isolines of a Gaussian field.  The nodal lines represent a sector of turbulence that can be described by an equilibrium statistical theory.

In this paper,  we aim to explain these results by proposing a novel treatment of the inverse cascade in 2d active scalar models.  The idea is that the inertial range is characterized by a spontaneous breaking of time reversal invariance, where a certain field acquires a non-zero expectation value.  In the case of Euler-Navier-Stokes hydrodynamics, the field is the product of the vorticity scalar and the fluid kinetic energy
\begin{align}
{\cal O}(x)  =  \frac{1}{2} \hat{\omega} \hat{v}^2.
\end{align}
Here the vorticity $\hat{\omega}$ and the velocity $\hat{v}^i$ are course-grained fluctuating fields in the turbulent flow.  ${\cal O}$ has the same dimensions as the energy flux, which are length per unit time cubed:
\begin{align}
[{\cal E}] =  [L] [T]^{-3}.
\end{align}
Ground states of the turbulent theory are associated with  ${\cal O}$ being equal to a constant, which yields the Kolmogorov-Kraichnan scaling of turbulent hydrodynamic fields by dimensional analysis.  We show that generalizations of this ``energy flux field'' $|{\cal O}|$ to hydrodynamic models with an active scalar field $\hat{\theta}$ produce the correct scaling of fields in all those cases.  In some active scalar models there is an enstrophy flux ${\cal A}$ in the inverse cascade instead of the energy flux.  In these examples, the nodal lines and their associated domains can be associated with the stream function $\hat{\psi}$.

We propose that the inverse cascade is associated with a phase transition where the order parameter takes different constant values in different regions of space.
The fully developed turbulent state consists of a random collection of domains separated by domain walls between the different ground states, which have positive and negative values of $\hat{\omega}$ or $\hat{\psi}$.  Like domain wall solutions in field theory, the field ${\cal O}$ interpolates between the states, with the domain wall located at the nodal lines of $\hat{\omega}$ or $\hat{\psi}$.

The turbulent ground states with different signs of $\hat{\omega}$ or $\hat{\psi}$ are distinguished by the generalized circulation of each region, e.g.
\begin{align}
\hat{\Gamma} = \int_V \hat{\omega} ~d^2 x.
\end{align}
Expressing the relevant scalar field in terms of the canonical Clebsch variables $(\mathbf{\beta}, \mathbf{\gamma})$ gives
\begin{align}
\hat{\Gamma} = \oint_C \mathbf{\beta} \partial_i \mathbf{\gamma}~ dx^i.
\end{align}
If $\mathbf{\gamma}$ is compact scalar (and $\mathbf{\beta}$ is a constant on the boundary $C$), then the generalized circulation of a domain is proportional to its winding number.  The winding number jumps across the nodal line boundaries.

Topology plays a crucial role in a variety of physical systems, particularly when time reversal invariance is broken.  Our picture of the inverse cascade in terms of a  dissipationless 2+1 dimensional system with a spontaneous breaking of time reversal is reminiscent of quantum Hall systems.  Some of our motivation comes from previous work \cite{Eling:2023iyx} where we showed that the ideal active scalar equations can be encoded into the equations of motion for a 2+1 dimensional abelian gauge theory with a Chern-Simons term.  Chern-Simons theory is the prototypical topological field theory, describing, for example, the low energy physics of quantum Hall systems (see, for example, \cite{Tong:2016kpv}).  The presence of a Chern-Simons term in the action indicates there is a topological sector in active scalar hydrodynamics, consisting of the zero energy degrees of freedom.  Presumably these degrees of freedom are also relevant in the turbulent inverse cascade.

Motivated by this analogy we propose a model of the turbulent domains in terms of constituent ``vortices'' with a minimal circulation at the forcing scale.  These vortices are analogous to the generically strongly coupled quasi-particles in the Hall fluid.  The nodal line domain boundaries where the local flux vanishes are associated with gapless topological degrees of freedom.

As a concrete model, we consider a 2d theory of the Clebsch scalars in the fully developed turbulent steady state.  In terms of the complex Clebsch scalar $\mathbf{\Psi} = \sqrt{\beta} e^{i \mathbf{\gamma}}$ the action is
\begin{align}
S_{2d} = \int d^2 x ~ \frac{1}{\pi g_0} \partial_\mu \mathbf{\Psi} \partial^\mu \mathbf{\Psi}^{\dagger}.
\end{align}
This theory has patch solutions, which are regions of $\hat{\theta}$ or $\hat{\psi}$ surrounded by $\hat{\theta}=0$ or $\hat{\psi}=0$. Inside these patches, hydrodynamic fields exhibit Kolmogorov-Kraichnan scaling.  Each patch has a constituent point vortex in the Clebsch field $\mathbf{\gamma}$ which generally has a fractional charge.  We argue that the fully developed inverse cascade is self-similar gas of these patches.  Furthermore, in this regime the Clebsch vortices form a critical state which is a Liouville CFT describing the gapless degrees of freedom.  The nodal lines are defect lines in this CFT.   The central charge of this CFT is determined by the Kolmogorov-Kraichnan scaling dimensions, which is in turn fixed by the charge of the vortices.  Our results are consistent with \cite{conformal1, conformal2, Falkovich}.

The organization of this paper is as follows.  Section 2 describes the phenomenology of turbulence in 2d active scalar models.  We show that the constant flux states of the inverse cascade can be described in terms of the order parameter of time reversal symmetry breaking, up to possible corrections due to the presence of the domain walls.  In Section 3 we consider the analogy between the inverse cascade and quantum Hall systems, focusing on the real space topology associated with the generalized circulation of the $\hat{\theta}$ or $\hat{\psi}$ domains.  Section 4 describes the theory of the Clebsch scalars in turbulence, the patch solutions, and our picture of the inverse cascade as an instanton gas.  Section 5 first reviews SLE theory and its connection to 2d CFT.  We then show how the central charge of the CFT can be fixed by the Kolmogorov-Kraichan scaling dimensions.  Finally, we conclude with a summary of results and discussion of some open problems.

\section{Turbulence and Inverse Cascade in 2d}

\subsection{Active Scalar Models}
\label{activescalars}

To start, we consider a scalar field $\theta(t,x)$ that obeys the following hydrodynamical evolution equation
\begin{align}
\partial_t \theta + \epsilon^{ij} \partial_j \psi \partial_i \theta = \eta \nabla^2 \theta - \alpha \theta + f.  \label{scalarhydroforcing}
\end{align}
$\psi(t,x)$ is the stream function associated with the divergence-free velocity field $v^i = \epsilon^{ij} \partial_j \psi$ that transports $\theta$.   $\eta$ is the viscosity and the $\alpha \theta$ term models a large-scale drag.  Finally, $f(\vec{x},t)$ represents an external forcing term acting on the system.

We also need the relationship between $\psi$ and $\theta$, which will subject the scalar to active transport. This relationship is given in general (formally) by the fractional Laplace equation
\begin{align}
\theta = -\Box^{m/2} \psi.  \label{fluidrelation}
\end{align}
In terms of Fourier space variables
\begin{align}
\theta_k = k^m \psi_k,
\end{align}
where $k = |\bf{k}|$ and $\bf{k}$ is the 2d wave number vector.  Different fluid models are described by different values of the parameter $m$.  $m=2$ is the Euler-Navier-Stokes system with $\theta$ equal to the fluid vorticity $\omega = \epsilon^{ij} \partial_i v_j  = -\Box \psi$.  Another value of interest is $m=1$, which is the surface quasi-geostrophic (SQG) model \cite{Held}.  In this case $\theta$ is not the vorticity, but instead the potential temperature.  Negative values of $m$ are also of physical relevance.  For example, $m=-2$ describes large scale flows of a rotating shallow fluid (the Charney-Obhukov-Hasegawa-Mima model) \cite{McWilliams1} in the limit of vanishing Rossby radius.

If the forcing term is taken as a Gaussian random variable, then (\ref{scalarhydroforcing}) becomes a complicated stochastic differential equation that can be used to model turbulent states.  The resulting statistics of the random $\hat{\theta}$ and $\hat{\psi}$ fields are not generally Gaussian since the stochastic equation is non-linear.

Suppose that energy is injected into the system by forcing term, which occurs on a characteristic length scale $L_f$.  Viscous dissipation of energy occurs on the smaller scale $L_d$.  For scales $L$  where $L_d \ll  L  \ll L_f$ (ie, where $\nu \rightarrow 0$) and $L_f \ll L \ll L_\alpha$ ($L_\alpha$ is associated with uniform friction) the idea is that the statistics of turbulent flows will have universal behavior in these inertial ranges, independent of the details of the forcing and the dissipation.

In the absence of forcing and dissipation there are an infinite number of conserved quantities associated with the ideal hydrodynamic equation (the left-hand side of (\ref{scalarhydroforcing})).  Kraichnan \cite{Kraichnan} argued that out of this infinity of conservation laws the two most relevant are quadratic positive definite forms, which for general $m$ are the ``energy'' $E$ and the enstrophy $A$:
\begin{align}
E = \frac{1}{2} \int d^2 x ~\psi \theta \nonumber \\
A = \frac{1}{2} \int d^2 x ~\theta^2
\end{align}
If there is a statistical steady state (ie, time independent statistics) then the existence of two conserved quantities implies then there are two turbulent cascades, each described by constant fluxes of energy or enstrophy. The direction of the fluxes is determined by the sign of the parameter $m$.

If $m > 0$ then there is a direct cascade of enstrophy to small scales where it is dissipated by viscosity.  Concurrently, energy undergoes an inverse cascade to larger scales ($L \gg L_f$) where it can be removed by large scale friction $\alpha$.  However, when $m < 0$, the cascades are reversed: enstrophy is transferred to large scales while energy cascades to small scales.

Here we will focus only on the inverse cascades.  Following Kolmogorov (see, eg. \cite{Pierrehumbert94}), one can assume scale invariance and analyze the problem in terms of dimensional analysis.  In the case of $m > 0$, the relevant quantity is the energy flux ${\cal E}$, the energy per unit volume per time.  In terms of the dimensions of $\psi$ and $\theta$
\begin{align}
[{\cal E}] = [T]^{-1} [\psi] [\theta].
\end{align}
The dimensions of $v$ and $\psi$ are always $[L] [T]^{-1}$ and $[L]^2 [T]^{-1}$, respectively. Hence, after taking spatial derivatives
\begin{align}
[{\cal E}] = [T]^{-3} [L]^{4-m} \label{Escaling}
\end{align}
To determine the scaling dimensions of the fields, we can assume that ${\cal E}$ is a dimensionless variable.  Hence, time and space should obey the following non-relativistic scaling relation
\begin{align}
T \sim L^{(4-m)/3}.
\end{align}
Therefore, in the inverse cascade, the stream function, the velocity and $\hat{\theta}$ should have the following scaling dimensions
\begin{align}
\hat{\psi} \sim L^{(2+m)/3} \nonumber \\
\hat{v} \sim L^{(m-1)/3} \nonumber \\
\hat{\theta} \sim L^{(2-2m)/3}. \label{positivescaling}
\end{align}
As we will see later, it is also interesting to define the dual velocity $\hat{\tilde{v}}^i = \epsilon^{ij} \partial_j \hat{\theta}$, which scales like
\begin{align}
\hat{\tilde{v}} \sim L^{(-2m-1)/3}.
\end{align}
For $m=2$, $\hat{\theta} \sim L^{-2/3}$ (Kolmogorov scaling) and for $m=1$, $\langle \theta(x') \theta(x) \rangle \sim \log(x-x')$.  Numerical simulations of the 2d inverse cascade have shown good agreement with these predictions for the scaling of correlation functions.

Note that restoring the proper dimensions of the fields brings in a (constant) energy flux factor ${\cal E}_0^{1/3}$ in (\ref{positivescaling}), eg,
\begin{align}
\hat{v} \sim {\cal E}_0^{1/3} L^{(m-1)/3}.
\end{align}

For $m < 0$ the enstrophy flux ${\cal A}$ has dimensions:
\begin{align}
[{\cal A}] = [T]^{-1} [\theta]^2  \label{Ascaling}
\end{align}
Following the same procedure as above, but now requiring a dimensionless enstrophy flux, yields:
\begin{align}
T \sim L^{2(2-m)/3}
\end{align}
In this case the scaling laws of the variables are
\begin{align}
\hat{\psi} \sim L^{(2+2m)/3} \\
\hat{v} \sim L^{(2m-1)/3} \\
\hat{\theta} \sim L^{(2-m)/3} \\
\hat{\tilde{v}} \sim L^{(-m-1)/3}.
\end{align}
If we restore the proper dimensions of the fields in this case, the formulas above pick up a factor of ${\cal A}_0^{1/3}$ similar to the $m>0$ case.

Note that there is a duality between the scaling dimensions of $\hat{\psi}$ and $\hat{\theta}$ and $\hat{v}$ and $\hat{\tilde{v}}$ when the sign of $m$ is flipped, ie
\begin{align}
m \rightarrow -m; ~~~ \hat{\theta} \rightarrow \hat{\psi} ~~ \hat{v} \rightarrow \hat{\tilde{v}}  \label{duality}.
\end{align}

\subsection{2d Inverse Cascade and Time Reversal Symmetry Breaking}
\label{trbreaking}

With the basic Kolmogorov-Kraichnan picture established, we now consider an alternative picture of the inverse cascade based on the spontaneous breaking of time reversal symmetry.  The spontaneous breaking appears when a field acquires a non-zero expectation value.  For the $m>0$ case, we take this field to be the product of $\hat{\theta}$ and $\hat{v}^2$:
\begin{align}
{\cal O}(x)_{m >0} =  \frac{1}{2} \hat{\theta} \hat{v}^2
\end{align}
The kinetic energy density $\sim \hat{v}^2$ is weighted by $\hat{\theta}$, which acts as a fluctuating rate; for $m=2$ the dimensions of $\theta$ are inverse time, $[T]^{-1}$.    For general $m$, the dimensions of ${\cal O}$ are the same as the dimensions of the energy flux ${\cal E}$.  Hence the absolute value of ${\cal O}$ acts as a localized energy flux to large scales
\begin{align}
|{\cal O}(x)| = {\cal E}(x).
\end{align}
We assume that the inverse cascade is characterized by states where ${\cal O}$ is constant
\begin{align}
{\cal O} = \pm {\cal O}_0,
\end{align}
${\cal O}$ is odd under time reversal and parity, so both discrete symmetries are spontaneously broken.  Furthermore, in these states the energy flux is also constant
\begin{align}
|{\cal O}_0| = {\cal E}_0.
\end{align}
In the turbulent ground states the velocity, stream function, $\theta$ fields exhibit the Kolmogorov-Kraichnan scaling, as discussed earlier.  The idea is that there is not a unique state with a constant energy flux, instead the turbulent theory has distinct states of $\pm {\cal O}_0$ where the energy flux is constant.  Note that in addition to the turbulent ground states, there is also in principle a regime associated with non-turbulent equilibrium states, where $\langle {\cal E}(x) \rangle = 0$ because there is no net energy flux.

As the fluid system is pumped by a forcing term, the system is driven towards one of the turbulent ground states.  However, when the forcing is random and acts on small scales, a soup of  $\pm {\cal O}_0$ patches begins to form.  Instead of settling into one ground state, the random forcing acts as disorder that creates domains of $+{\cal O}_0$ and $-{\cal O}_0$ separated by domain wall configurations.  These domain wall solutions interpolate between the different ground states, for example ${\cal O}$ typically has the characteristic domain wall profile
\begin{align}
{\cal O} = {\cal O}_0 \tanh \left(\frac{x-r_0}{W} \right),
\end{align}
where $r_0$ is the location of the wall and $W$ is its width.  When the steady state is reached, the inverse cascade that is created consists of a set of domains separated by random $\hat{\theta} = 0$ domain walls.

As we will discuss in more detail in the next section, the crucial point is that these domains of constant energy flux are distinguished by different values of a topologically invariant winding number. Furthermore, there are topologically protected ``gapless'' degrees of freedom living on the domain walls, where the localized energy flux vanishes. The gapless nature of the degrees of freedom on the wall suggests the domain walls exhibit a conformal symmetry.  We will argue later that in the turbulent steady state, the walls are equivalent to critical curves in a conformal field theory, as observed in numerical simulations.

In the case where $m<0$, we want to find a similar order parameter with the same dimensions as the enstrophy flux.  A quantity that fits this requirement is
\begin{align}
{\cal O}(x)_{m < 0} = \frac{1}{2} \hat{\psi} \hat{\tilde{v}}^2.
\end{align}
Note that under the duality discussed earlier (\ref{duality}), ${\cal O}_{m<0}$  can be mapped into ${\cal O}_{m>0}$
\begin{align}
m \rightarrow -m;   ~~~~  {\cal O}_{m >0}  \rightarrow {\cal O}_{m<0}
\end{align}
The duality between the field and associated order parameters explains the duality of the scaling dimensions in the inverse cascade, which was obscure in the standard picture.

The ground states with ${\cal O}_{m <0}  = \pm {\cal O}_0$ are equivalent to states with constant enstrophy flux that exhibit the $m<0$ Kolmogorov-Kraichnan scaling.  The arguments are the same as for $m>0$, but now we expect the inverse enstrophy cascade consists of domains of $\hat{\psi}$ separated by random domain walls of $\hat{\psi} = 0$, where the local enstrophy flux ${\cal A}(x)$ vanishes.

In $m<0$ inverse cascade it was observed that $\hat{\psi} = 0$ nodal lines are SLE curves, like the $\hat{\theta}= 0$ when $m>0$ \cite{Falkovich}.  This special behavior of $\hat{\psi}$ nodal curves was puzzling because the stream function appears to have no special physical significance in the active scalar models.  For example, it is not a Lagrangian invariant of the ideal flow, unlike $\theta$.  Here we will see that the $\hat{\psi} = 0$ curves are special in the inverse cascade because they act as domain walls separating topologically distinct states of constant enstrophy flux.

Finally, our treatment of the energy and enstrophy fluxes in terms of domain wall configurations raises the question whether the Kolmogorov-Kraichnan scaling of the fields could be subject to corrections.  In the direct cascade of 3d turbulence there are the intermittent corrections we mentioned earlier.  For example, Kolmogorov and Obukhov \cite{K62} proposed a modified version of the 1941 theory where the energy flux is a varying field.  In this case the structure functions for the longitudinal velocity increments $\delta \hat{v}^{\parallel}(r)$ take the form
\begin{align}
\langle (\delta \hat{v}^\parallel)^n \rangle = \langle {\cal E}_\ell^{n/3} \rangle  r^{n (m-1)/3}
\end{align}
${\cal E}_\ell$ is an average of the local energy flux over some region of size $\ell$ and is treated as a fluctuating field with its own statistics.  In the 3d direct cascade ${\cal E}_\ell$ is expected to be highly intermittent, subject to large fluctuations.  Higher powers of $n$ are influenced by rare events.  The energy flux is expected to be concentrated on a small part of the spatial volume, at filaments or sheets of intense vorticity.   The average flux is constant, $\langle {\cal E}_\ell \rangle = constant$, consistent with the exact result that $\langle (\delta \hat{v})^3 \rangle \sim r$, but higher moments of the velocity increments do not scale like $n(m-1)/3$.

In our case, there is also a fluctuating energy flux.  However, while vortex filaments are associated with intense dissipation, our domain walls are regions of weak dissipation.  The domain walls apparently lead to exponentially suppressed corrections, e.g.
\begin{align}
{\cal E}(x) = {\cal E}_0 \left(1 - e^{-x/W} + \dots \right),
\end{align}
where $x$ is the distance from the wall.  If we assume that the correlation length $\xi$ of the fluctuations of ${\cal E}(x)$ is small on the scale of the inverse cascade $\xi \ll L$ (as we've argued the fluctuations are associated with small-scale disordering due to the forcing), then it appears that the mean field description of the energy flux is valid and any corrections to Kolmogorov-Kraichnan scaling are small.  On the other hand, the small corrections could sum to a significant contribution if there are large number of domain walls. For example, if the domain walls condense and proliferate throughout space, then the ordered constant flux states will break down and we do not expect Kolmogorov-Kraichnan scaling to be valid.

Interestingly, recent numerical work \cite{Mueller} has shown that while 2d longitudinal differences in the $m=2$ Euler case obey Kolmogorov-Kraichnan scaling, the statistics of transverse velocity differences $(\delta v)^\perp = \delta \vec{v} - \delta v^{\parallel} \hat{r}$ and moments of circulation $\Gamma_r$ about fixed loops $C_r$ of size $r$
\begin{align}
\Gamma_r = \oint_{C_r} v^i dx_i,
\end{align}
show intermittent deviations.  Transverse velocity differences are in general much more intermittent, for example in 3d turbulence it is known that the exponents $\zeta_n$ in $\langle (\delta v^\perp)^n \rangle \sim r^{\zeta_n}$ saturate at about 2 for $n>10$ \cite{Sreenivasan}. There is some evidence 3d longitudinal differences may saturate to a larger number at much higher $n$.  In 2d, the $\zeta_n$ do not show indications of saturation.  The presence of intermittency indicates there may be some strong fluctuations in ${\cal E}$ not captured by our theory.  Perhaps the intermittency in 2d becomes negligible as the idealized viscosity $\eta = 0$ regime is approached.  However, in the following we will be conservative and restrict attention to the scale invariant sector of 2d turbulent flows.

\section{Topology in the inverse cascade}
\label{topology}

We have postulated that the inverse cascade is comprised of degenerate states with constant fluxes of energy or enstrophy separated by domain walls where the local fluxes vanish.  The constant flux states break time reversal invariance and can be thought of as gapped states.  The energy ``gap'' $\delta E$ associated with a domain is equal to the energy flux ${\cal E}_0$ divided by the average circulation density $\hat{\Gamma}$ associated with the domains of positive or negative $\hat{\theta}$
\begin{align}
\Delta E = \frac{{\cal E}_0}{\frac{\hat{\Gamma}}{R^2}},
\end{align}
where $R$ is the characteristic radius of the domain.  An analogous expression can be written for the enstrophy gap in the $m<0$ case.

The constant flux states are reminiscent of well-known gapped 2d systems in condensed matter physics, such as quantum Hall systems \cite{Tong:2016kpv} or the more general class of Chern insulators.  In the quantum Hall effect, time reversal invariance is broken by an external magnetic field piercing the 2d system of electrons in the plane.  Over time it has been realized that time reversal can be broken intrinsically in certain lattice systems without an external magnetic field \cite{Haldane:1988zza,Neupert:2011fea,Sheng:2011gea,Regnault:2011zz}.  In continuum field theory, the classic example is a massive Dirac fermion in 2+1 dimensions, which is not invariant under time reversal or parity.   Our model of the constant flux states is similar to a 2d Dirac fermion with a spatially varying mass term that breaks time reversal (for example, a fermion coupled to a scalar field with a domain wall profile).  The mass is analogous to our ${\cal E}$ and ${\cal A}$.

What makes these condensed matter systems special is that although they are gapped, the low energy physics of hydrodynamic fluctuations and linear response is non-trivial due to topological effects, which require the existence of gapless edge modes at a system boundary.  For example, the eigenstates of the Hamiltonian are distinguished by different Chern winding numbers in momentum space (integrals of the Berry phase).

%Furthermore, the bulk-boundary correspondence [citation??] implies that the difference in bulk winding numbers is equal to the number of chiral fermion zero modes on the domain wall boundary.

In the case of the flux states we don't have an analog of the 2d Dirac Hamiltonian or a clear way to calculate linear response to external perturbations\footnote{The turbulent states could have, for example, a dissipationless Hall viscosity in response to fluctuations of the background metric}.  Do the turbulent states actually have topological properties? To proceed, we will consider a real space topological charge which is related to the generalized circulation of the domains of positive (or negative) $\hat{\theta}$ or $\hat{\psi}$.

We start by introducing Clebsch variables $(\mathbf{\beta},\mathbf{\gamma})$.  In terms of these variables $\hat{\theta}$ has the form
\begin{align}
\hat{\theta} = \epsilon^{ij} \partial_i \mathbf{\beta} \partial_j \mathbf{\gamma}. \label{Clebschdef}
\end{align}
For an ideal fluid, the Clebsch variables are Lagrangian invariants transported by the velocity $v^i$
\begin{align}
\partial_t \mathbf{\beta} + v^i \partial_i \mathbf{\beta} = 0 \nonumber \\
\partial_t \mathbf{\gamma} + v^i \partial_i \mathbf{\gamma} = 0.
\end{align}
In the Hamiltonian formulation of the ideal fluid equations, the Clebsch variables are the canonical coordinates $(p,q)$.

We define a topological current in 2+1 dimensions ($x^\mu = (x^0, x^i)$) as
\begin{align}
J^\mu = \frac{1}{\Gamma_0} \epsilon^{\mu \nu \rho} \partial_\nu \mathbf{\beta}  \partial_\rho \mathbf{\gamma}. \label{current2}
\end{align}
$\Gamma_0$ is some basic unit of circulation which we will discuss in more detail shortly.   The charge density has the form
\begin{align}
J^0 = \frac{1}{\Gamma_0} \hat{\theta}.
\end{align}
Integrating over a region $V$ yields the charge, which is proportional to the generalized circulation
\begin{align}
Q = \int_V d^2x ~J^0 = \frac{1}{\Gamma_0} \oint_C \mathbf{\beta} \partial_i \mathbf{\gamma} ~dx^i,
\end{align}
where $\partial V = C$ and we have used Stokes’ theorem to write the volume integration as a contour integral.

In the $m<0$ case we consider $\hat{\psi}$ and the dual circulation
\begin{align}
\tilde{Q} = \frac{1}{\tilde{\Gamma}_0} \int_V d^2 x  ~\hat{\psi}.
\end{align}
This doesn't look like a circulation, but when, for example, $m=-2$, $\hat{\psi} = \nabla^2 \hat{\theta}$. Then $\tilde{Q}$ can be re-expressed as a line integral over $\tilde{v}^i$
\begin{align}
\tilde{Q} = \frac{1}{\tilde{\Gamma}_0} \oint_C \tilde{v}^i dx_i
\end{align}
In this case we can introduce a dual set of Clebsch variables parametrizing $\hat{\psi}$ that is transported by $\tilde{v}^i$.  In the following we will typically work with the $m>0$ case, but all our results can easily be translated into their dual representation for $m<0$ with $\hat{\psi}$.

In the turbulent flow the clusters of $\hat{\theta}$ or $\hat{\psi}$ are random.  Hence the cluster boundary $C$ is also random. We assume the curve is fractal but has no self-intersections, i.e. it is the outer boundary of the cluster.  To treat this issue we consider a representative loop $C=C^\star$ from the ensemble and average over the hydrodynamic fields (e.g. Clebsch variables) such that $\hat{\theta} = 0$ or $\hat{\psi} = 0$ on $C^\star$.   The idea is that the averaged generalized circulation $\hat{Q}$ will be proportional to a topological number.

For this to be true, we must allow the Clebsch variable $\mathbf{\gamma}$ to be a compact multi-valued field that jumps by $2\pi$ around a closed loop. In 3d turbulence Migdal has recently proposed a theory where the Clebsch scalars have a similar behavior as compact variables and behave as confined fields in turbulence, analogous to quarks in QCD \cite{Migdal:2022bka}.  Since $\mathbf{\gamma}$ is compact, we could also equivalently work with a complex Clebsch scalar
\begin{align}
\mathbf{\Psi} = \sqrt{\mathbf{\beta}} e^{i \mathbf{\gamma}},
\end{align}
and its complex conjugate $\mathbf{\Psi}^\dag$, in analogy with the  Madelung representation of a superfluid wavefunction.  Second, $\mathbf{\beta} = \mathbf{\beta}_0$ must be a constant along the integration contour $C=C^{\star}$.  Then it follows that
\begin{align}
\hat{Q} = \frac{\mathbf{\beta}_0}{\Gamma_0} ~ \oint_{C^\star} \partial_i \mathbf{\gamma} dx^i = \frac{2\pi \nu \mathbf{\beta}_0}{\Gamma_0},
\end{align}
where $\nu$ is the winding number.  $\nu>0$ corresponds to positive circulation (cluster of positive $\hat{\theta}$) while $\nu < 0$ corresponds to negative circulation and negative $\hat{\theta}$.   For example, in the simplest case we can imagine a cluster domain of positive or negative $\hat{\theta}$ surrounded by a region of $\hat{\theta} = 0$. At the cluster boundary $\mathbf{\beta} = \mathbf{\beta}_0$ and $\mathbf{\beta}$ remains equal to $\mathbf{\beta}_0$ in the zero $\hat{\theta}$ region exterior of the cluster.

The inverse cascade consists of random clusters that fill the entire space.  Our picture is that the inverse cascade is a gas of non-overlapping clusters of positive and negative generalized circulation separated by regions where $\hat{\theta} = 0$ or $\hat{\psi} = 0$.  The circulation in each cluster is proportional to a winding number of a compact Clebsch variable.

$\mathbf{\beta}_0$ has dimensions of circulation and acts as the flux of the time reversal symmetry breaking field (analogous to the magnetic flux in the quantum Hall system).  If we insert the Kolmogorov-Kraichnan scaling for the circulation using (\ref{positivescaling}), we find
\begin{align}
\hat{Q} \sim \frac{2\pi \nu}{\Gamma_0} {\cal E}_0^{1/3} A_D^{\frac{4-m}{3}},
\end{align}
where $A_D$ is the area of the region/droplet. For $m<0$ a similar result follows for the case of the enstrophy flux in terms of the quantity $\int_V \hat{\psi} d^2 x$.

Since $\hat{Q}$ is a dimensionless variable, we can think of it as equal to (on average) the number of positive (or negative) ``constituents'' $N$ in the region.  In this picture $\nu$ is analogous to the filling fraction in the quantum Hall system
\begin{align}
\nu = \frac{N \Gamma_0}{2\pi \mathbf{\beta}_0}.
\end{align}
$\Gamma_0$ is the circulation of each constituent and $\mathbf{\beta}_0$ carries the information about the total circulation of the region.

In the quantum Hall fluid, the constituents are the electrons or quasi-particle excitations.  What are the microscopic constituents of the turbulent states?  The classical turbulence of ordinary fluids is not quantum mechanical and doesn't involve $\hbar$.  There is a special case where classical turbulence presumably emerges as the limit of superfluid turbulence, which involves many point vortices with quantized circulation $\Gamma_0 = 2\pi \hbar/m$, where $m$ is the particle mass.   A quantum Hall fluid can also be described as type of superfluid \cite{Stone:1990cd, Abanov:2012smr,  Monteiro:2022wip} with a ``Hall relation'' constraining vorticity to be equal to $n\Gamma_0/\nu$.   In the superfluid analogy, $ \beta= \mathbf{\Psi} \mathbf{\Psi}^\dag$ is the density of the Clebsch quasi-particles in the cluster.

We will assume that in the inverse cascade the turbulent flow is made up of many approximately point-like sources of $\hat{\theta}$ or $\hat{\psi}$ created by the small scale stirring of the system.  For example, a generalized stirring force $F$ acting on some small-scale $L_f$ and over a time $t_f$ creates tiny vortices with a minimal circulation
\begin{align}
\Gamma_0 \sim F \cdot L_f \cdot t_f.
\end{align}
To recap, our picture is that in the inverse cascade consists of a random soup of clusters (or droplets) of positive and negative generalized circulation. The clusters are composed of many tiny vortex constituents, analogous to the quasi-particles in a quantum Hall fluid.

At the domain wall cluster boundaries, the local energy flux vanishes and the topological winding number jumps.  We argue that both properties are crucial to the observed behavior of the $\hat{\theta} = 0$ and $\hat{\psi} = 0$ curves.  This scenario is roughly similar to a system of 2d Dirac fermions with a spatially varying, random mass term $\hat{M}(x,y)$.  The mass term corresponds to $\hat{\theta}$ or $\hat{\psi}$ in the sense that a change in the sign of these variables corresponds to a jump in a topological number.  In the fermion case, the Hall conductivity equals $\pm \frac{1}{2}$ in units of $e^2/2\pi \hbar$ depending on the sign of $M$ (this is true even in a pure system with non-random mass)\cite{Ludwig1994}.  In the inverse cascade, the winding number changes from $+\nu$ to $-\nu$ or vice versa across the $\hat{\theta}$ and $\hat{\psi}$ nodal lines.  The difference is that the random fluid variables are ``active'', following from the turbulent dynamics, while the random mass term is external and passive.

The fermion system with random mass is a simple model of a disordered Hall system, with $\hat{M}(x,y)$ corresponding to a random potential term $\hat{V}(x,y)$.  In quantum Hall physics, electrons move along equipotential lines of $\hat{V}(x,y)$.   Around the maxima and the minima of the potential, the electrons tend to be localized, unable to propagate through the sample.  However, along equipotential lines where $\hat{M}(x,y) = \hat{V}(x,y) = 0$ electrons are delocalized.  These correspond to the gapless edge states, which can percolate through the system \cite{ChalkerCoddington}.  Therefore, the transition between quantum Hall states is a localization-delocalization phase transition, with the localization length approaching infinity at the critical point.  The critical point is expected to be described by a conformal field theory.

In the turbulent case we argue that in the fully developed steady state there is a similar phase transition at the $\hat{\theta}$ or $\hat{\psi}$ nodal lines.  The nodal lines can be described as defect lines in a conformal field theory, which will be explored later.  One could heuristically argue that there is a similar localization-delocalization phenomenon involving the small-scale vortices that make up the inverse cascade, but this is somewhat speculative\footnote{For some evidence that vortices can have restricted mobility, see \cite{fractons}}.

In the following section we investigate a simple 2d model of the inverse cascade in terms of scalar patch configurations of a Clebsch scalar field theory.  This model realizes many of our conceptual arguments above and allows us to relate the Kolmogorov-Kraichnan scaling of the active scalar models to different winding numbers or ``filling fractions'' of each cluster.  A Liouville CFT describes the gapless, conformal degrees of the freedom in the system, which explains the SLE curves and the central charges found in \cite{conformal1,conformal2,Falkovich}.

\section{Scalar patch model of the Inverse Cascade}
\label{Instantons}

The domains of positive and negative $\hat{\theta}$ or $\hat{\psi}$ with different winding numbers that we introduced above are reminiscent of instanton configurations in a field theory.  In the steady state, the turbulent flow is effectively two-dimensional, so we consider an effective description of the system in terms of a 2d Euclidean field theory of the Clebsch scalars.  The simplest 2d action for a complex Clebsch scalar $\mathbf{\Psi}$ is
\begin{align}
S_{2d} = \int d^2 x ~ \frac{1}{\pi g_0} \partial_\mu \mathbf{\Psi} \partial^\mu \mathbf{\Psi}^{\dagger},  \label{Clebschaction}
\end{align}
where $g_0$ is a coupling.  We could also include a topological (total derivative) term with parameter $\bar{\theta}$
\begin{align}
S_{\rm top} = \int d^2x \frac{\bar{\theta}}{2\pi} \epsilon^{\mu \nu} \partial_\mu \mathbf{\Psi} \partial_\nu \mathbf{\Psi}^{\dagger}
\end{align}
This term is proportional to the generalized circulation density and breaks time reversal and parity explicitly if there is a net circulation in the system. We will neglect this term and assume the net circulation is zero.

The action (\ref{Clebschaction}) is invariant under $U(1)$ phase rotations of $\mathbf{\Psi}$, so it is similar to the $O(2)$ XY model for a superfluid.  However, the usual additional $U(1)$ invariant potential terms are absent because we also want symmetry under the shifts $\mathbf{\Psi} \rightarrow \mathbf{\Psi} + a$.  The action above can be thought of as the effective description of a kind of ``Clebsch superfluid'' state.  The real Clebsch variables $\mathbf{\beta}$ and $\mathbf{\gamma}$ defined above in (\ref{Clebschdef}) represent the superfluid density and the phase, respectively.

In terms of the real Clebsch variables, the action has the form
\begin{align}
S_{2d} = \int d^2 x ~ \frac{1}{\pi g_0} \left( \frac{1}{4 \mathbf{\beta}} \partial_\mu \mathbf{\beta} \partial^\mu \mathbf{\beta} + \mathbf{\beta} \partial^\mu \mathbf{\gamma} \partial_\mu \mathbf{\gamma} \right).
\end{align}
The next step is to notice that this action can be rewritten in the Bogomolnyi form familiar from instanton physics, where there is a squared term plus the topological term which is the generalized circulation above
\begin{align}
S_{2d} = \int d^2 x ~ \frac{1}{\pi g_0} \left(\frac{\partial_\mu \beta}{2 \sqrt{\mathbf{\beta}}} \pm \sqrt{\mathbf{\beta}} \epsilon_\mu{}^\nu \partial_\nu \mathbf{\gamma}\right)^2 \mp \int d^2 x~  \frac{1}{\pi g_0} \epsilon^{\mu \nu} \partial_\mu \beta \partial_\nu \mathbf{\gamma}.
\end{align}
Hence, if the fields satisfy the self-duality equation
\begin{align}
\frac{\partial_\mu \mathbf{\beta}}{2 \sqrt{\mathbf{\beta}}} \pm \sqrt{\mathbf{\beta}} \epsilon_\mu{}^\nu \partial_\nu \mathbf{\gamma} = 0,
\end{align}
then the action can be minimized to be equal to the topological circulation charge.  Note that we can re-express the duality equation in more of a gauge theoretical form as
\begin{align}
\frac{1}{2} \partial_\mu \ln \mathbf{\beta} \pm \epsilon_\mu{}^\nu H_\nu,
\end{align}
where the second term is the Hodge dual of one-form field $H_\mu = \partial_\mu \mathbf{\gamma}$.  The action is similar to a (two-dimensional) version of a scalar dilaton field coupled to a one-form gauge field \cite{Rey:1989xj}. The self-duality equation expresses a balance between the field strengths of the two fields.

We will first solve the duality equation using a spherically symmetric ansatz. Because $\mathbf{\gamma}$ is a phase, we can take $\mathbf{\gamma} = \nu \phi$ for polar coordinate $\phi$ and winding number $\nu$, which will initially be an integer. The duality equation reduces to
\begin{align}
\partial_r \ln \mathbf{\beta}(r) = \pm \frac{2 \nu}{r},
\end{align}
where $r$ is the radial coordinate.  The solution of this equation is simply
\begin{align}
\mathbf{\beta} = C r^{\pm 2 \nu}.
\end{align}
for some constant $C$.

A drawback of these solutions is that they are divergent in the IR (or UV) and don't approach a constant at infinity, which is needed for the topological charge to be finite.  However, for $r^{2\nu}$ we can cut off the solution at some radius $r = R$, creating a circular droplet of radius $R$, and demand that for $r \geq R$, $\mathbf{\beta} = \mathbf{\beta}_0 = C R^{2\nu}$.  For $\nu = 1$ the solution is a patch of constant positive $\hat{\theta}$ surrounded by a region of $\hat{\theta} = 0$. We will refer to these as scalar patches. For $\nu = -1$ the $\hat{\theta}$ is negative and we take the anti-self dual solution to keep the IR divergent $\mathbf{\beta}$.

This piecewise solution is technically no longer an instanton because the constant $r \geq R$ piece does not satisfy the self-dual equation.  Inserting the solution into the action yields
\begin{align}
S_{2d} =  \frac{2 C R^{2 |\nu|}}{g_0} \left(|\nu| + \nu^2 \log \frac{L_c}{R}\right),  \label{actionvalue}
\end{align}
where $L_c$ is some cutoff scale representing the size of the system.  Our scalar patch solution has some resemblance to the meron solutions discussed in Yang-Mills gauge theories and non-linear sigma models \cite{Callan:1977qs,Callan:1977gz,Gross:1977wu}, which are piecewise and also have logarithmically divergent actions.  The logarithmic term here is the Coulomb interaction associated with the point vortex in $\mathbf{\gamma}$ located at the origin.

Before we continue further, there is a slight generalization which connects the scalar patch solutions more directly with the turbulent inverse cascade.  Let's return to the choice of $\mathbf{\gamma} = \nu \phi$ in the construction of the scalar patch solution.  Suppose that we analytically continue the winding number to be a rational number and not necessarily an integer.  If $\nu$ is a rational number, then under $\phi \rightarrow \phi + 2\pi$, the complex Clebsch variable picks up an overall phase factor
\begin{align}
\mathbf{\Psi} \rightarrow e^{2 \pi i \nu} \mathbf{\Psi}
\end{align}
This is not necessarily a problem since the phase of the Clebsch ``wavefunction'' is not observable in the actual velocity and other fields characterizing a fluid flow.  We could interpret the phase as an anyonic Aharanov-Bohm factor that appears when a point vortex is transported around a loop surrounding the patch.  As in the previous section, we'll think of the $\mathbf{\Psi}$ as a mean field in the turbulent flow, not a classical field of a smooth flow.  The effective fractional topological charge arises from the complicated strongly coupled dynamics of the constituent vortices.

Moreover, when $\nu$ is a rational number the scaling of $\mathbf{\beta}$ can be matched to the Kolmogorov-Kraichnan scaling of turbulent fields.  Because $\mathbf{\gamma}$ is dimensionless, (\ref{Clebschdef}) implies that the scaling of $\hat{\theta}$ in the patch is
\begin{align}
\hat{\theta} = \epsilon^{ij} \partial_i \mathbf{\beta} \partial_j \mathbf{\gamma} \sim L^{2|\nu| - 2}
\end{align}
In addition, the generalized circulation scales like the area of the region to the power $\nu$,
\begin{align}
\Gamma_C = 2 \pi^{1-|\nu|} |\nu| {\cal E}_0^{1/3} A^{|\nu|}_D,
\end{align}
where we have chosen the constant $C$ to be the energy flux, consistent with a spontaneous breaking of time reversal symmetry. Matching to Kolmogorov-Kraichnan scaling fixes $\nu$ to be
\begin{align}
|\nu| = \frac{4-m}{3}.  \label{chargemapping}
\end{align}

Our picture is that the fully developed turbulent state consists of many of these fractionally charged patches with different winding numbers $\pm \nu$.  Therefore, we need to consider general multi-patch solutions.  In general, if we take the curl and divergence of the self-duality equation we find
\begin{align}
\nabla^2 \gamma =& 0  \\
\frac{1}{2} \nabla^2 \log \mathbf{\beta} =& \sum_i \nu_i \delta^2(\vec{x}-\vec{X}_i).
\end{align}
(Point) vortices in $\mathbf{\gamma}$ are located at $\vec{X}_i$.  Each vortex is balanced by the field due to $\log \mathbf{\beta}$, creating the patch of topological charge density dependent on the patch variable $\vec{R}_i$.  In a multi-patch solution, we have patches with size $|\vec{R}_i - \vec{X}_i|$ separated by regions where $\mathbf{\beta} = \mathbf{\beta}_0$ takes a uniform value.  Because $\mathbf{\beta}_0$ is determined by the size of the patch, this means that the sizes of the patches should be the same.

Following the discussion in the previous section, we propose that near the forcing scale, there are many patches of size $R \sim L_f$ separated by regions of $\hat{\theta} = 0$ or $\hat{\psi} = 0$.  Inside the patches the fields have a scale invariant power law form. Suppose we have a tiny loop encircling a constituent vortex (effectively a tiny patch) carrying a fractional charge/circulation. One could also imagine a lattice of size $R$ with sites where $\nu$ can be zero or plus/minus a generally fractional value.  At each site on the dual lattice $\mathbf{\beta}$ scales like $R^{2|\nu|}$.  As we go to larger scales, the constituents renormalize into larger vortex patches.  As we perform a course graining process over the vortices and scale $R' = R/\lambda$, the large-scale coursed grained fields describing the turbulent flow (e.g. $\hat{\theta}$ or $\hat{\psi}$) maintain their Kolmogorov-Kraichnan form, at least in the mean.

In our quantum Hall analogy in the previous section, $\nu$ is the filling fraction of each patch region.   When $m=1$, $|\nu| = 1$ and the number of vortex constituents scales like the area (volume) of the region, completely filling the region in some sense.  In the Euler case of $m=2$, $|\nu| = 2/3$ and the number of constituents scales like $A^{2/3}_D$.  We can also consider some other cases of general $m$.  For example, for $m=1/2$, $|\nu| = 5/6$, while for $m=5/2$, $|\nu| = 1/2$.

We can also think about our picture of the turbulent flow in the context of statistical field theory.  The partition function for our effective theory (\ref{Clebschaction}) can be expressed as the grand canonical partition function for the gas of patches.  The logarithmic term in (\ref{actionvalue}) is associated with regions where $\hat{\theta}=0$ or $\hat{\psi}=0$ and arises due to the Coulomb interaction between Clebsch vortices.  Moreover, (\ref{actionvalue}) is similar to the form of the free energy of a point vortex, e.g,
\begin{align}
F_{\rm vortex} = A q^2 \log \frac{L_c}{a} + F_{\rm core},
\end{align}
where $A$ is some coupling, $q$ is the charge, $a$ is a UV cutoff.  Usually, $F_{\rm core}$ represents an energy due to the core of the vortex below the cutoff, which acts as a chemical potential.  Here the topological term in (\ref{actionvalue}) acts like the energy from the interior patch ``core'' and depends on the size of the patches.

In our case $q=\nu$, and $a$ is identified with the radius $R$ of the patch.  Furthermore, we can also identify a size dependent coupling $g$
\begin{align}
\frac{2}{g} = \frac{C R^{2 |\nu|}}{g_0}
\end{align}
In terms of the coupling $g$, $F_{\rm core} = 2|\nu|/g$.  Suppose that we consider patches of a small, fixed size much smaller than their separations.  In the partition function we would integrate over the locations of the vortices and anti-vortices, which we assume is the same so that system is neutral, and the total action is finite.

In the Coulomb gas system, vortices are confined or deconfined depending on the effective temperature, which here is related to $g$.  It is tempting to connect the presence of vortices in the Clebsch field $\mathbf{\gamma}$ with the onset of turbulence in the fluid flow.  At low temperatures Clebsch vortices are confined into vortex-antivortex pairs.  This regime corresponds to smooth fluid flows with a few scalar patches.  However, as the temperature increases, a phase transition occurs and the Clebsch vortices then become unbounded and proliferate, forming a plasma phase.  In this regime, the patches dissolve or ``melt'', which corresponds to incoherent $\hat{\theta}$ or $\hat{\psi}$.  At the critical point, which corresponds to the fully developed turbulent state, the Clebsch vortex gas is scale invariant.

This phase transition scenario mirrors (in reverse order) the process of decaying turbulence seen in numerical simulations (see e.g., \cite{McWilliams}).  The initial state consists of an incoherent $\hat{\theta}$ corresponding to the deconfined Clebsch vortices produced by stirring at small scales.  The Clebsch vortices then become more confined as time increases, corresponding to the condensation of patches and emergence of the fully developed turbulent state exhibiting scale invariance.  The final state at late times consists of a smooth flow with a small number of vortex patches.  Here the Clebsch vortices are confined.

Typically, temperature (and pressure) are external parameters of the system.  Phase transitions occur when a system is heated (or cooled) or the pressure is changed.  Instead, the fluid system seems to realize the Clebsch vortex phase transition through its own dynamics, which would change $g$. This picture of the turbulent flow is reminiscent of self-organized criticality \cite{Bak}, where the variety of fractal and power law behavior observed in nature follows from non-linear, non-equilibrium systems self-tuning to critical states. Note however that the critical fully developed steady state does require a degree of fine-tuning because it can only be maintained by the small-scale force continuously injecting energy into the system.

To recap, we propose that the state of fully developed turbulence in the inverse cascade is a scale invariant gas of scalar patches. In this picture, the $\hat{\theta}$ and $\hat{\psi}$ nodal lines are boundaries of these patches, where the scale invariance is enhanced to conformal invariance.  We propose that these curves are defect lines in the Coulomb gas of Clebsch vortices at criticality.  In the following section we will discuss this idea in more detail, connecting the behavior of nodal lines as SLE curves with the critical Coulomb gas as a Liouville CFT.

\section{SLE curves and a Liouville CFT description of topological degrees of freedom in the Inverse Cascade}
\label{Liouville}

We now discuss the behavior of the $\hat{\theta} =0$ and $\hat{\psi} = 0$ lines in fully developed turbulence.  Our aim is to provide a theoretical explanation of the results of \cite{conformal1, conformal2, Falkovich}, where $\hat{\theta}=0$ or $\hat{\psi} = 0$ are different types of SLE curves, depending on the choice of the parameter $m$.

First, we briefly review SLE curves and their connection to conformal field theory.  SLE curves are random paths taken from boundary point to boundary point on a region of the complex plane. There are several varieties of SLE, but here the relevant case will be chordial SLE, which describes random paths taken from boundary point to boundary point on a region of the complex plane.    One typically considers the upper half plane $\bf{H}$ with the real axis as the boundary.  SLE describes the evolution of a conformal mapping $g_\lambda(z)$ that takes the simply connected region outside of the curve with a ``slit'' geometry $\gamma(\lambda)$, ie ${\bf H} \backslash \gamma(\lambda)$, back to $\bf{H}$.  The tip of $\gamma$ is mapped back to an image point on the real axis $\xi(\lambda)$.  Conversely, $\xi(\lambda)$ is a driving function on the real axis and the inverse conformal map maps this into the tip of the growing curve in $\bf{H}$.  For each tip point $\lambda$ there is a different conformal map.  The evolution of the mapping under $\lambda$ is described by Loewner's equation, with the following asymptotic behavior that fixes the choice of ``hydrodynamic'' time parametrization:
\begin{align}
\frac{d g_\lambda(z)}{d\lambda} = \frac{2}{g_\lambda(z)-\xi(\lambda)} \nonumber \\
g_\lambda(z) \sim z + 2\lambda/z + O(1/z^2).
\end{align}
Schramm discovered that if $\xi(\lambda) = \sqrt{\kappa} B_\lambda$, where $B_\lambda$ is a normalized one-dimensional Brownian motion and $\kappa$ is a dimensionless diffusion constant, then the resulting random curves obey a Markov property and have a conformally invariant measure \cite{Schramm:1999rp}.

The wiggliness of a SLE curve depends on the value of $\kappa$. For $0 \leq \kappa \leq 4$, $\gamma$ is simple and does not intersect itself or the real axis, ie, it escapes to infinity. For $4 < \kappa < 8$ the curve intersects itself and the real axis, creating closed hull regions of $\bf{H} \backslash \gamma$.  The outer perimeter of the hull turns out to be a SLE curve with parameter $\kappa'$ given by the duality relation
\begin{align}
\kappa’ = 16/\kappa
\end{align}
For $\kappa \geq 8$ the curve fills the entire space.

In \cite{conformal1, conformal2, Falkovich} the authors performed numerical analyses of the inverse cascade in different active scalar models. Using an arbitrarily chosen real axis, they found sets of $\hat{\theta} = 0$ curves emanating from the axis.  It is then possible to map (approximately) the curves back to the real axis following the Loewner procedure.  The question is whether the associated driving function is Brownian.

Remarkably, $\xi(\lambda)$ is a Brownian walk and the value of $\kappa$ can be determined from Gaussian statistics of $\xi$ .  For the $m=2$ Euler case $\kappa = 6$ (and $\kappa' = 8/3$ for outer cluster boundaries). $\kappa = 4$ in the $m=1$ SQG case.   Interestingly, for $0<m \leq 1$, \cite{Falkovich} found that the SLE curves still have $\kappa = 4$ and $c=1$.  Finally, in the $m=-2$ case the value of $\kappa=6$ describes $\hat{\psi}=0$ isolines, consistent with the $m \rightarrow -m$ duality described above in Section \ref{activescalars}.

The conformal properties of SLE suggest that there should be a direct link to 2d CFT.  The link can be established in the context of boundary CFT, ie, a CFT in the presence of a boundary, with boundary conditions consistent with conformal symmetry \cite{DiFrancesco:1997nk}.  In the CFT picture a boundary condition changing operator (e.g. an operator associated a change from a spin up configuration to spin down configuration on the real axis) creates a defect line that can be identified with an SLE curve, yielding the following relation between diffusion constant $\kappa$ and the central charge $c$ \cite{Bauer:2006gr}
\begin{align}
c = \frac{(3\kappa-8)( \kappa-6)}{2\kappa}. \label{centralcharge1}
\end{align}
For $\kappa = 6$ (and $\kappa' = 8/3$) the corresponding central charge is zero. For $\kappa = 4$, $c= 1$.

We have argued that the $\hat{\theta} =0$ and $\hat{\psi} = 0$ curves are avatars of gapless topological degrees of freedom in the turbulent flow, forming the boundaries between the random clusters of $\pm \nu$ in seen in numerical simulations. In the previous section we proposed that in the turbulent state there is a Coulomb gas of Clebsch vortices.  The simplest case with charges $\pm 1$ can be mapped into a sine-Gordon theory
\begin{align}
S_{\rm sg} = \int d^2 x  \left(\frac{1}{8\pi} \nabla_\mu \mathbf{\Phi} \nabla^\mu \mathbf{\Phi} + \lambda (e^{i \frac{\sqrt{2}}{\sqrt{g}}  \mathbf{\Phi}} + e^{-i \frac{\sqrt{2}}{\sqrt{g}}  \mathbf{\Phi}})\right),
\end{align}
In the special case of $g = 1$ the operators $e^{\pm i \frac{\sqrt{2}}{\sqrt{g}} \mathbf{\Phi}}$ are marginal, with conformal weight 1.  The theory is conformal at this value, with central charge $c=1$.

However, in general the turbulent state involves fractional charge $\nu$.  To describe the theory at the critical point we first propose that $g=|\nu|$, so that the effective temperature or coupling of the system at criticality is equal to the charge.  However, for general values of $\nu<1 $, the operator $e^{i \frac{\sqrt{2}}{\sqrt{|\nu|}} \mathbf{\Phi}}$ cannot be made marginal if $c=1$. The solution is to couple the theory to a background scalar curvature $R$, which shifts the weight of vertex operators.

We propose that the conformal sector of the inverse cascade can be described by a Liouville CFT with the following action
\begin{align}
S_{\rm Liouville} = \int d^2 x \left( \frac{1}{8\pi} \nabla_\mu \mathbf{\Phi} \nabla^\mu \mathbf{\Phi} + i \frac{1-|\nu|}{4\pi \sqrt{2|\nu|}} R \mathbf{\Phi} +  \mu e^{i \frac{\sqrt{2}}{\sqrt{|\nu|}} \mathbf{\Phi}} \right) + i \int dx \frac{1-|\nu|}{2\pi \sqrt{2|\nu|} } K \mathbf{\Phi}.
\end{align}
$K$ the extrinsic curvature of a boundary \cite{Kondev:1997dy, Bettelheim:2005mt, criticalcurves}.  The coefficient in front of the curvature terms is chosen so that the Liouville potential term is a marginal operator.  In the Coulomb gas picture of CFT \cite{DiFrancesco:1997nk} this is related to defining a screening charge operator with dimension zero
\begin{align}
Q_{\rm screening} = \int d^2 x ~e^{i \frac{\sqrt{2}}{\sqrt{|\nu|}} \mathbf{\Phi}}.
\end{align}
This theory is called an imaginary compact Liouville theory \cite{Guillarmou:2023exh}, due to the imaginary action\footnote{In the general parametrization of the Liouville potential $e^{b \mathbf{\Phi}}$ the coupling $b$ is imaginary.} and the compact scalar field.  The imaginary Liouville theory is expected to describe the continuum limit of statistical loop models.  The central charge of the theory is modified by the coupling to the curvature
\begin{align}
c = 1 - 6 \frac{(1-|\nu|)^2}{|\nu|}.  \label{centralnu}
\end{align}
In the turbulent steady state the $\pm \nu$ regions are associated with the chiral and anti-chiral halves of the Liouville CFT.  These chiral theories describe the gapless, topological degrees of freedom in the turbulent flow.  The conformal blocks in those regions can be interpreted as being analogous to wavefunctions for the vortex constituents in the limit of a large number of constituents.  The two sectors (left and right) can be glued together at the boundary.  We can then use the ``folding trick'' \cite{Oshikawa:1996dj} to fold the anti-chiral (antiholomorphic) degrees of freedom to the left side, creating a full CFT with a random boundary.

On the patch boundary surfaces, where $\hat{\theta} = 0$ or $\hat{\psi} = 0$  and the winding number changes sign, the compact boson $\mathbf{\Phi}$ experiences a jump associated with the presence of a defect line.  SLE curves are defect lines created by a particular degenerate field $V_{12} = e^{-i \frac{\sqrt{2}}{\sqrt{|\nu|}} \mathbf{\phi}}$ at the boundary associated with a null state at level 2 in the Virasoro algebra.

%This vertex operator has a magnetic charge that is associated with a vortex in the field $\mathbf{\Phi}$.  As one makes a complete circle around the vortex there is a discontinuity line in the field.

%In the fully turbulent picture, the locations of the fluctuating boundaries can be interpreted in terms of a fluctuating metric \cite{Bauer:2002tf}.  In the purely 2d picture of the steady state, the low energy effective action depends only on the 2d metric.  The form of the effective action is determined by conformal anomaly, which famously leads to the Liouville action \cite{Polyakov}.

Comparing the formula for $c$ in terms of $\nu$ above (\ref{centralnu}) to the earlier relation for $c$ in terms of $\kappa$ (\ref{centralcharge1}), yields the relationship between $\nu$ and the SLE diffusion constant: $|\nu| = 4/\kappa = \kappa'/4$.  Furthermore, the relation (\ref{chargemapping}) between $|\nu|$ and $m$ we found in Section \ref{Instantons} via the turbulent scaling dimensions yields following relationship between $\kappa$ and $m$ for $m>1$
\begin{align}
\kappa =& \frac{12}{4-m} \nonumber \\
\kappa' =& \frac{4}{3} \left(4-m\right),
\end{align}
which agrees with the proposal in \cite{Falkovich}.  For the Euler case $|\nu| = 2/3$, $\kappa = 6$, $\kappa' = 8/3$ and $c=0$, which was found originally in \cite{conformal1}.  Note that when $|\nu| = p/q$ for integer $p$ and $q$, the central charge takes the form of the Virasoro minimal models when $p$ and $q$ are coprime
\begin{align}
c = 1 - 6 \frac{(q-p)^2}{pq}
\end{align}
In particular, $p=2$ and $q=3$ correspond to the $c=0$ theory.  $p=5$ and $q=6$ yield $c=4/5$, which corresponds to $m=3/2$.  For $q = p+1$, the filling fractions are $\nu = p/(p+1)$.

In cases where the parameter $5/2 \leq  m < 4$, the filling fraction $\nu \leq 1/2$, $\kappa \geq 8$, and $c \leq -2$.  In this case, the SLE curves fill the entire plane and have a fractal dimension equal to 2.  It appears that the entire picture of cluster boundaries and domains breaks down in this regime.  Furthermore, note that $\hat{\theta} \sim L^{-x}$ for $x \geq 1$ so that $\hat{\theta}$ is strongly dependent on the small-scale physics of the constituent vortices, which corresponds to the domain wall curves becoming more and more wiggly and eventually space filling.

The remaining case is when $0 < m < 1$.  In these cases, it has been observed that the $\hat{\theta} = 0$ lines have $\kappa = 4$ and thus are defect lines in a CFT with $c=1$.   In our picture, this implies that the filling fraction/charge $|\nu| = 1$ for $0<m \leq 1$ and the ``level'' appears to be saturated
\begin{align}
|\nu| = \frac{4-m}{3}, ~~~~ m > 1 \nonumber \\
|\nu| = 1,  ~~~~ 0 < m \leq 1
\end{align}
The behavior of the filling fraction also appears to be connected with the scaling exponents for $\hat{v}$ and $\hat{\theta}$ in (\ref{positivescaling}), which change sign at $m=1$.  When $0<m \leq 1$, $\hat{v}$ has a negative exponent and $\hat{\theta}$ has a positive exponent, which means that the velocity is a small-scale field while $\hat{\theta}$ is a large-scale field.  Recall that the integer quantum Hall effect follows when interactions between the electrons can be neglected.  In our case this corresponds to neglecting the interactions between the small-scale vortex patch constituents of the turbulent flow.  Indeed, heuristically, if $\hat{\theta}$ is a large-scale field, then the small scale vortices have negligible contribution.

In contrast, for $m > 1$ $\hat{v}$ is a large-scale field and $\hat{\theta}$ is small-scale.  This corresponds to the case where the small-scale behavior of the vortices is important, which leads to the more complicated physics of fractional charges.  Furthermore, in this regime the velocity differences grow with distance and affect the IR physics.

Finally, we note that a similar Liouville theory was considered by Vafa in \cite{Vafa:2015euh} as a part of a proposal to realize the fractional quantum Hall effect in the context of M-theory.  In this case the marginal Liouville potential is parametrized as $e^{i \frac{1}{\sqrt{\nu'}}\mathbf{\Phi}}$ so that the minimal model CFT's are associated with the two principal series of filling fractions seen experimentally in the quantum Hall effect.  The filling fraction $\nu'$ is related to our $\nu$ via $2\nu' = |\nu|$.  It was also pointed out that the Liouville theory is dual to 2+1 dimensional Chern-Simons theory with $SL(2, \mathbb{R}) \times SL(2, \mathbb{R})$ gauge group in Minkowski space or to $SL(2, \mathbb{C})$ in 3d Euclidean space.  Hence, the low-energy effective description of the inverse cascade appears to be encoded into this type of Chern-Simons theory, which is closely related to gravity in 3 dimensions \cite{Witten:1988hc}.

\section{Discussion}

In this paper we have developed an alternative picture of hydrodynamic turbulence of 2d active scalar models, motivated by the observation that the nodal lines of some scalar fields are SLE curves and display conformal invariance in the fully developed inverse cascade.  First, we argued that the inverse cascade is associated with a certain operator acquiring a non-zero expectation value, which leads to the spontaneous breaking of time reversal invariance.  The breaking of this discrete symmetry leads to the formation of domain wall configurations separating states with different generalized circulations.  In cases where the energy is transferred to larger scales, the domain walls are located at the nodal lines of the active scalar field $\hat{\theta}$.  However, in cases where the enstrophy flux is inverse, the domain walls are located at nodal lines of the stream function $\hat{\psi}$.  Away from the domain walls, the local energy and enstrophy fluxes are constant.  This yields the Kolmogorov-Kraichnan scaling of the fluctuating hydrodynamic fields up to possible small corrections.

In this approach, turbulent states with constant fluxes of energy (or enstrophy) are distinguished by opposite values of a topological winding number $\nu$ associated with opposite signs of the generalized circulation.  In particular, the turbulent flow consists of connected regions or domains where $\hat{\theta}$ or $\hat{\psi}$ are proportional to positive or negative values $\pm \nu$.  To model this behavior we introduced Clebsch variables to parametrize $\hat{\theta}$ and $\hat{\psi}$.   At the nodal line cluster boundaries, where the local fluxes are zero, the scale invariance of the system is enhanced to full 2d conformal invariance.  The nodal lines can be thought of as slices of equilibrium physics in the turbulent state.  Working in analogy with quantum Hall fluids, we argued heuristically that the turbulent flow consists of a large number of ``vortex'' patch constituents with a minimal circulation fixed by forcing scale.

We then introduced a simple two-dimensional effective theory of the turbulent system which realizes many aspects of this picture. The Clebsch scalars comprise the modulus and phase of a complex scalar field.  The action for complex Clebsch scalar is invariant under $U(1)$ phase rotations and shift translations.  This theory has solutions that spontaneously break time reversal invariance and correspond to finite size patches of $\hat{\theta}$ or $\hat{\psi}$.  Inside these patches the hydrodynamic fields exhibit Kolmogorov-Kraichnan scaling if the winding number associated with the point vortex in the phase of the complex Clebsch field is fractional.  We argued that the scale invariant sector of the turbulent flow can be viewed as an instanton gas of these fractionally charged patches.

Finally, we proposed the $\hat{\theta}$ or $\hat{\psi}$ nodal lines are defect lines in a Liouville CFT that describes the critical point of the fractionally charged Clebsch vortex gas. The central charge of the CFT is determined by the winding number/charge, which is in turn related to the Kolmogorov-Kraichnan scaling exponents of the turbulent theory. Our results agreed with the formula relating the central charge and the active scalar model proposed earlier in \cite{Falkovich}.  Note that in our picture the conformal invariance is hidden in the gapless degrees of freedom associated with the nodal lines. The turbulent hydrodynamic fields are gapped and have Kolmogorov-Kraichan scaling dimensions. Therefore, these are not the primary operators of the Liouville CFT.

In the future it would be interesting to connect our approach more directly with the active scalar equations, for example via the gauge theory mapping in \cite{Eling:2023iyx}.  It also would be useful to flesh out more details of the scalar patch gas approach.  This likely will involve going beyond the mean field regime and considering fluctuations of the patches.  Conformal invariant nodal lines have also been detected in other types of systems beyond the active scalar models (see, e.g. \cite{optical}).  Perhaps our picture can be generalized to these cases.

Finally, there is also the question of whether (or to what degree) these ideas can be generalized to direct cascades, both in 2d and 3d turbulence, where significant intermittency is seen in all types of statistics. Most of our methods have been specialized to 2d and the inverse cascade, but it may be possible to understand the 3d turbulence as a similar type of vortex cell gas (see, e.g. \cite{Migdal:1993bg}).


\begin{thebibliography} {99}

\bibitem{Frischbook} U.~Frisch, ``Turbulence: the legacy of A.~N.~Kolmogorov,'' Cambridge university press, 1995.

\bibitem{Kraichnan} R.~H.~Kraichnan, ``Inertial ranges in two-dimensional turbulence,''
Physics of fluids \textbf{10}, no. 7 1417 (1967).

\bibitem{Mueller}
N.~P.~M\"{u}ller and G.~Krstulovic, ``Lack of self-similarity in transverse velocity increments and circulation statistics in two-dimensional turbulence,'' Phys. Rev. Fluids \textbf{10} (1), L012601 (2025).

\bibitem{DiFrancesco:1997nk}
P.~Di Francesco, P.~Mathieu and D.~Senechal,
``Conformal Field Theory,''
Springer-Verlag, 1997,
ISBN 978-0-387-94785-3, 978-1-4612-7475-9
doi:10.1007/978-1-4612-2256-9

\bibitem{Polyakov:1992er}
A.~M.~Polyakov,
``The Theory of turbulence in two-dimensions,''
Nucl. Phys. B \textbf{396}, 367-385 (1993)
doi:10.1016/0550-3213(93)90656-A
[arXiv:hep-th/9212145 [hep-th]].

\bibitem{Falkovich:1993fg}
G.~Falkovich and A.~Hanany,
``Is 2-d turbulence a conformal turbulence?,''
Phys. Rev. Lett. \textbf{71}, 3454-3457 (1993)
doi:10.1103/PhysRevLett.71.3454
[arXiv:hep-th/9301030 [hep-th]].

\bibitem{conformal1} D.~Bernard, G.~Boffetta, A.~Celani, and G.~Falkovich, ``Conformal invariance in two-dimensional turbulence,'' 
Nature Physics~ \textbf{2}, 2, 124 (2006) [arxiv: nlin/0602017 [nlin.CD]].

\bibitem{Schramm:1999rp}
O.~Schramm,
``Scaling limits of loop-erased random walks and uniform spanning trees,''
Isr. J. Math. \textbf{118}, 221-288 (2000)
doi:10.1007/BF02803524
[arXiv:math/9904022 [math.PR]].

\bibitem{Bauer:2006gr}
M.~Bauer and D.~Bernard,
``2D growth processes: SLE and Loewner chains,''
Phys. Rept. \textbf{432}, 115-221 (2006)
doi:10.1016/j.physrep.2006.06.002
[arXiv:math-ph/0602049 [math-ph]].

\bibitem{conformal2} D.~Bernard, G.~Boffetta, A.~Celani, and G.~Falkovich, ``Inverse turbulent cascades and conformally invariant curves,''
Phys. Rev. Lett.~ \textbf{98}, 2 (2007) 024501 [arxiv: nlin/0609069 [nlin.CD]].

\bibitem{Falkovich} G.~Falkovich and S.~Musacchio,
``Conformal invariance in inverse turbulent cascades'' 
[arXiv:1012.3868 [cond-mat.stat-mech]].

\bibitem{FalkovichRev}
G.~Falkovich, ``Conformal invariance in hydrodynamic turbulence,''
Russian Mathematical Surveys \textbf{62}, no. 3 497 (2007).

\bibitem{Eling:2023iyx}
C.~Eling,
``A gauge theory for the 2+1 dimensional incompressible Euler equations,''
[arXiv:2305.04394 [hep-th]].

\bibitem{Tong:2016kpv}
D.~Tong,
``Lectures on the Quantum Hall Effect,''
[arXiv:1606.06687 [hep-th]].

\bibitem{Held}  I.~M.~Held, R.~T.~Pierrehumbert, S.~T.~Garner, and K.~L.~Swanson,
``Surface quasi-geostrophic dynamics,''
J. Fluid Mech. 282 (1995) 1-20.

\bibitem{McWilliams1} V.~Larichev and J.~McWilliams, ``Weakly decaying turbulence in an equivalent barotropic fluid,''
Phys. Fluids, A 3:5 (1991) 938.

\bibitem{Pierrehumbert94}
R.~T.~Pierrehumbert, I.~M.~Held, and K.~L.~Swanson, ``Spectra of local and nonlocal two-dimensional turbulence,''
Chaos, Solitons and Fractals \textbf{4}, no. 6 (1994): 1111-1116.

\bibitem{K62}
A.~N.~Kolmogorov,  ``A refinement of previous hypotheses concerning the local structure of turbulence in a viscous incompressible fluid at high Reynolds number,'' J. Fluid Mech., \textbf{13} 82, (1962);  A.~M.~Obukhov, ``Some specific features of atmospheric turbulence,'' J. Fluid Mech., \textbf{13} 77 (1962).

\bibitem{Sreenivasan}
K.~P.~Iyer, K.~R.~Sreenivasan, and P.~K.~Yeung, ``Scaling exponents saturate in three-dimensional isotropic turbulence,''
Physical Review Fluids \textbf{5}, 054605 (2020).


\bibitem{Haldane:1988zza}
F.~D.~M.~Haldane,
``Model for a Quantum Hall Effect without Landau Levels: Condensed-Matter Realization of the 'Parity Anomaly',''
Phys. Rev. Lett. \textbf{61}, 2015-2018 (1988)
doi:10.1103/PhysRevLett.61.2015

\bibitem{Neupert:2011fea}
T.~Neupert, L.~Santos, C.~Chamon and C.~Mudry,
``Fractional Quantum Hall States at Zero Magnetic Field,''
Phys. Rev. Lett. \textbf{106},  23, 236804 (2011)
doi:10.1103/PhysRevLett.106.236804
[arXiv:1012.4723 [cond-mat.str-el]].


\bibitem{Sheng:2011gea}
D.~N.~Sheng, Z.~C.~Gu, K.~Sun and L.~Sheng,
``Fractional quantum Hall effect in the absence of Landau levels,''
Nature Communications \textbf{2}, 389 (2011)
doi:10.1038/ncomms1380
[arXiv:1102.2658 [cond-mat.str-el]].

\bibitem{Regnault:2011zz}
N.~Regnault and B.~A.~Bernevig,
``Fractional Chern insulator,''
Phys. Rev. X \textbf{1}, 021014 (2011)
doi:10.1103/PhysRevX.1.021014
[arXiv:1105.4867 [cond-mat.str-el]].

%\bibitem{Callan:1984sa}
%C.~G.~Callan, Jr. and J.~A.~Harvey,
%``Anomalies and Fermion Zero Modes on Strings and Domain Walls,''
%Nucl. Phys. B \textbf{250}, 427-436 (1985)
%doi:10.1016/0550-3213(85)90489-4


%
%\bibitem{Wiegmann:2013hca}
%P.~Wiegmann and A.~G.~Abanov,
%``Anomalous Hydrodynamics of Two-Dimensional Vortex Fluids,''
%Phys. Rev. Lett. \textbf{113}, no.3, 034501 (2014)
%doi:10.1103/PhysRevLett.113.034501
%[arXiv:1311.4479 [physics.flu-dyn]].

\bibitem{Migdal:2022bka}
A.~Migdal,
``Statistical equilibrium of circulating fluids,''
Phys. Rept. \textbf{1011}, 1-117 (2023)
doi:10.1016/j.physrep.2023.02.001
[arXiv:2209.12312 [physics.flu-dyn]].


\bibitem{Stone:1990cd}
M.~Stone,
``SUPERFLUID DYNAMICS OF THE FRACTIONAL QUANTUM HALL STATE,''
Phys. Rev. B \textbf{42}, no.1, 212 (1990)
doi:10.1103/PhysRevB.42.212

\bibitem{Abanov:2012smr}
A.~G.~Abanov,
``On the effective hydrodynamics of the fractional quantum Hall effect,''
J. Phys. A \textbf{46}, 292001 (2013)
doi:10.1088/1751-8113/46/29/292001
[arXiv:1212.0461 [cond-mat.str-el]].

\bibitem{Monteiro:2022wip}
G.~M.~Monteiro, V.~P.~Nair and S.~Ganeshan,
``Topological fluids with boundaries and fractional quantum Hall edge dynamics: A fluid dynamics derivation of the chiral boson action,''
Phys. Rev. B \textbf{109}, no.17, 174525 (2024)
doi:10.1103/PhysRevB.109.174525
[arXiv:2203.06516 [cond-mat.mes-hall]].

\bibitem{Ludwig1994}
A.~W.~W~Ludwig, M.~P.~A.~Fisher, R.~Shankar, and G.~Grinstein, ``Integer quantum Hall transition: An alternative approach and exact results.''
Physical Review B \textbf{50}, no. 11, 7526 (1994).

\bibitem{fractons}
D.~Doshi and A.~Gromov, ``Vortices as fractons.''
Communications Physics \textbf{4}, no. 1, 44 (2021).

\bibitem{ChalkerCoddington}
J.~T.~Chalker and P.~D.~Coddington, ``Percolation, quantum tunnelling and the integer Hall effect.''
Journal of Physics C: Solid State Physics \textbf{21}, no. 14, 2665 (1988).

\bibitem{Rey:1989xj}
S.~J.~Rey,
``The Confining Phase of Superstrings and Axionic Strings,''
Phys. Rev. D \textbf{43}, 526-538 (1991)
doi:10.1103/PhysRevD.43.526


\bibitem{Callan:1977qs}
C.~G.~Callan, Jr., R.~F.~Dashen and D.~J.~Gross,
``A Mechanism for Quark Confinement,''
Phys. Lett. B \textbf{66}, 375-381 (1977)
doi:10.1016/0370-2693(77)90019-3

\bibitem{Callan:1977gz}
C.~G.~Callan, Jr., R.~F.~Dashen and D.~J.~Gross,
``Toward a Theory of the Strong Interactions,''
Phys. Rev. D \textbf{17}, 2717 (1978)
doi:10.1103/PhysRevD.17.2717

\bibitem{Gross:1977wu}
D.~J.~Gross,
``Meron Configurations in the Two-Dimensional O(3) Sigma Model,''
Nucl. Phys. B \textbf{132}, 439-456 (1978)
doi:10.1016/0550-3213(78)90470-4


\bibitem{McWilliams}
J.~M.~McWilliams, ``The emergence of isolated coherent vortices in turbulent flow,''
J. Fluid Mech. \textbf{146} 21 (1984)

\bibitem{Bak}
P.~Bak, C.~Tang, and K.~Wisenfeld,
 ``Self-organized criticality: an explanation of $1/f$ ~noise,''
 Phys. Rev. Lett \textbf{59}, 381 (1987).

\bibitem{Kondev:1997dy}
J.~Kondev,
``Liouville field theory of fluctuating loops,''
Phys. Rev. Lett. \textbf{78}, 4320-4323 (1997)
doi:10.1103/PhysRevLett.78.4320
[arXiv:cond-mat/9703113 [cond-mat]].

\bibitem{Bettelheim:2005mt}
E.~Bettelheim, I.~Rushkin, I.~A.~Gruzberg and P.~Wiegmann,
``On harmonic measure of critical curves,''
Phys. Rev. Lett. \textbf{95}, 170602 (2005)
doi:10.1103/PhysRevLett.95.170602
[arXiv:hep-th/0507115 [hep-th]].

\bibitem{criticalcurves}
I.~Rushkin, E.~Bettelheim, I.~ A.~Gruzberg, and P.~Wiegmann,
``Critical curves in conformally invariant statistical systems,''
 Journal of Physics A: Mathematical and Theoretical \textbf{40}, 9, 2165 (2007).

\bibitem{Guillarmou:2023exh}
C.~Guillarmou, A.~Kupiainen and R.~Rhodes,
``Compactified Imaginary Liouville Theory,''
[arXiv:2310.18226 [math-ph]].

%\bibitem{Bauer:2002tf}
%M.~Bauer and D.~Bernard,
%``Conformal field theories of stochastic Loewner evolutions,''
%Commun. Math. Phys. \textbf{239}, 493-521 (2003)
%doi:10.1007/s00220-003-0881-x
%[arXiv:hep-th/0210015 [hep-th]].


\bibitem{Oshikawa:1996dj}
M.~Oshikawa and I.~Affleck,
``Boundary conformal field theory approach to the critical two-dimensional Ising model with a defect line,''
Nucl. Phys. B \textbf{495}, 533-582 (1997)
doi:10.1016/S0550-3213(97)00219-8
[arXiv:cond-mat/9612187 [cond-mat]].

\bibitem{Vafa:2015euh}
C.~Vafa,
``Fractional quantum Hall effect and $M$-theory,''
Adv. Theor. Math. Phys. \textbf{27}, no.1, 1-36 (2023)
doi:10.4310/ATMP.2023.v27.n1.a1
[arXiv:1511.03372 [cond-mat.mes-hall]].

\bibitem{Witten:1988hc}
E.~Witten,
``(2+1)-Dimensional Gravity as an Exactly Soluble System,''
Nucl. Phys. B \textbf{311}, 46 (1988)
doi:10.1016/0550-3213(88)90143-5.

\bibitem{optical}
R.~Panico,  A.~S.~Lanotte, D.~Trypogeorgos, G.~Gigli, M.~De Giorgi, D.~Sanvitto, and D.~Ballarini,
``Conformal invariance of 2D quantum turbulence in an exciton-polariton fluid of light,''
 Applied Physics Reviews \textbf{10}, no. 4 (2023).

\bibitem{Migdal:1993bg}
A.~A.~Migdal, ``Turbulence as statistics of vortex cells,''
[arXiv:hep-th/9306152 [hep-th]].


\end{thebibliography}
\end{document}